\documentclass[12pt]{article}
\usepackage{amsmath, amssymb}
\usepackage{graphicx}
\newcommand{\etal}{\textit{et al. }}
\newcommand{\la}{\langle}
\newcommand{\ra}{\rangle}
\newcommand{\cell}[1]{\makebox[0.4cm]{#1}}

\begin{document}

\title{Critical behavior of a cellular automaton highway traffic model}

\author{Nino Boccara\dag\ and Henryk Fuk\'s\ddag\\
\dag\ Department of Physics, University of Illinois, Chicago,
USA\\ \texttt{boccara@uic.edu}\\ and\\ DRECAM/SPEC, CE Saclay,
91191 Gif-sur-Yvette Cedex, France \vspace{0.4cm} \\ \ddag\ The
Fields Institute for Research in Mathematical Sciences,\\ Toronto
ON M5T 2W1, Canada\\ \texttt{hfuks@fields.utoronto.ca}\\ and\\
Department of Mathematics and Statistics, University of Guelph, \\
Guelph, ON N1G 2W1, Canada}

\maketitle

\noindent \textit{Keywords}: cellular automata, highway traffic,
second-order phase transitions, critical phenomena.

\vspace{0.5cm}

{\small{\textbf{Abstract.} We derive the critical behavior of a CA
traffic flow model using an order parameter breaking the symmetry
of the jam-free phase. Random braking appears to be the
symmetry-breaking field conjugate to the order parameter. For
$v_{\max}=2$, we determine the values of the critical exponents
$\beta$, $\gamma$ and $\delta$ using an order-3 cluster
approximation and computer simulations. These critical exponents
satisfy a scaling relation, which can be derived assuming that the
order parameter is a generalized homogeneous function of
$|\rho-\rho_c|$ and $p$ in the vicinity of the phase transition
point.}}

\vspace{0.5cm}

\section{Introduction}
Cellular automaton (CA) models of traffic flow have attracted much
interest since the publication of the Nagel-Schreckenberg (NS)
model \cite{NaSch}. The NS model is a probabilistic CA model of
traffic flow on a one-lane highway. The road is represented by a
lattice of L sites with periodic boundary conditions. Each site is
either empty (in state $e$) or occupied by a car with velocity
$v=0,1,2,\ldots,v_{\max}$ (in state $v$). If $d_i$ is the distance
between car $i$ and car $i+1$ (cars are moving to the right),
velocities are updated in parallel according to the following two
subrules:
\begin{align}
v_i(t+1/2) & = \min(v_i(t)+1, d_i(t)-1, v_{\max})\label{R1}\\
v_i(t+1) & = \max(v_i(t+1/2)- 1, 0)\quad \text{with
probability}\;\;p,\label{R2}
\end{align}
where $v_i(t)$ is the velocity of car $i$ at time $t$; then, cars
are moving according to the subrule:
\begin{equation}
x_i(t+1) = x_i(t) +  v_i(t+1),\label{R3}
\end{equation}
where $x_i(t)$ is the position of car $i$ at time $t$. The model
contains three parameters: the maximum speed $v_{\max}$, which is
the same for all cars, the braking probability $p$, and the car
density $\rho$. For a clear presentation of the various
approximate techniques to calculate relevant physical quantities
of the NS model compared with results obtained from computer
simulations see \cite{Schad}.

If $p=0$, the NS model is deterministic and the average velocity
over the whole lattice is exactly given by
\begin{equation}
\la v\ra = \min\left(v_{\max}, \frac{1}{\rho} - 1\right).
\label{avspeed}
\end{equation}
This expression shows that, below a critical car density $\rho_c =
1/(v_{\max}+1)$, all cars move with a velocity equal to
$v_{\max}$, while above $\rho_c$, the average velocity is less
than $v_{\max}$. This transition from a free-moving regime to a
congested regime is usually viewed as a second-order phase
transition.

If $p\neq 0$, except in the case $v_{\max}=1$, no exact expression
for the average velocity has been obtained, and there is no
consensus concerning the existence of the phase transition (see,
for instance \cite{ESSS} and \cite{LSU} and references therein).
One point that is particularly unclear is the definition of the
order parameter.

If CA models of traffic flow exhibit second-order phase
transitions, we have to understand the nature of the order
parameter, show how it is related to symmetry-breaking, determine
the symmetry-breaking field conjugate to the order parameter,
define the analogue of the susceptibility, study the critical
behavior, and, possibly, find scaling laws.

The purpose of this short note is to fulfill this program. In our
discussion we shall consider a slightly simpler model of traffic
flow. We shall assume that the acceleration, which is equal to 1
in the NS model, has the largest possible value (less or equal
to $v_{\max}$) as in the Fukui-Ishibashi (FI) model \cite{FK},
except that these authors apply random
delays only to cars whose velocity is equal to $v_{\max}$. That
is, in our model, we just replace (\ref{R1}) by
\begin{equation}
v_i(t+1/2) = \min(d_i(t)-1, v_{\max})\label{R1'}.
\end{equation}

\section{Symmetry considerations}
Deterministic CA rules modeling traffic flow on one-lane highways
are number-conserving ($\rho=\text{constant}$). Limit sets of
number-conserving CA have, in most cases, a very simple structure
and, these limit sets are reached after a number of time steps
proportional to the lattice size \cite{BNR, BF1, BF2}. The limit set
of our model is identical to the limit set of the NS or FI models.

If $\rho<\rho_c$, any configuration in the limit set consists of
``perfect tiles'' of $v_{\max} + 1$ cells as shown below
\begin{center}
\begin{tabular}{|c|c|c|c|c|}
\hline
\cell{$e$}&\cell{$e$}&\cell{$\cdots$}&\cell{$e$}&\cell{$v_{\max}$}\\
\hline
\end{tabular}
\end{center}
in a sea of cells in state $e$. In particular, it can be shown
\cite{HF} that the probability to find in the limit set a sequence
of $v_{\max}+1$ empty sites is exactly given by
$$
P(\underbrace{ee\cdots e}_{v_{\max}}) = 1 - (v_{\max}+1)\rho.
$$
This relation is a simple consequence of the following argument:
If, in an empty road, you add $L\rho$ cars, in the limit set,
for $\rho<\rho_c$, we have
$$
L P(\underbrace{ee\cdots e}_{v_{\max}}) = L - (v_{\max}+1)L\rho,
$$
where $L$ is the lattice size.

If $\rho>\rho_c$, a configuration belonging to the limit set
only consists of a mixture of tiles containing $v+1$ cells of
the type
\begin{center}
\begin{tabular}{|c|c|c|c|c|}
\hline
\cell{$e$}&\cell{$e$}&\cell{$\cdots$}&\cell{$e$}&\cell{$v$}\\
\hline
\end{tabular}
\end{center}
where $v = 0,1,\cdots,v_{\max}$.
If $\{\rho_v\mid v=0,1,2,\cdots,v_{\max}\}$ is the velocities
distribution, we have
\begin{align*}
\rho & = \sum_{v=0}^{v_{\max}}\rho_v\\
1 & = \sum_{v=0}^{v_{\max}}(v+1)\rho_v\\
\la v\ra & = \frac{1}{\rho}
\left(\sum_{v=0}^{v_{\max}}v\rho_v\right).
\end{align*}
Note that (\ref{avspeed}) is a simple consequence of these
relations.

If we introduce random braking, then, even at low density, some
tiles become defective, which causes the average velocity to be
less than $v_{\max}$. The random-braking parameter $p$ can,
therefore, be viewed as a symmetry-breaking field, and the order
parameter, conjugate to that field is
\begin{equation}
m = v_{\max} - \la v\ra.\label{op}
\end{equation}
This point of view implies that the phase transition characterized
by $m$ will be smeared out in the presence of random braking as in
ferromagnetic systems placed in a magnetic field.

From (\ref{op}) and (\ref{avspeed}), it follows that, for $p=0$,
\begin{equation}
m =
\begin{cases}
0 & \text{if $\rho\leq\rho_c$},\\ [0.5em] \displaystyle
\frac{\rho-\rho_c}{\rho\rho_c} & \text{otherwise}.
\end{cases}
\end{equation}
Hence, the critical exponent $\beta$ is equal to 1.

\section{Approximate techniques}
To determine the other critical exponents, we have used
local structure approximations as described in [10] and numerical
simulations.

To construct a local structure approximation, it is more
convenient to represent configurations of cars as binary
sequences, where zeros represent empty spaces and ones represent
cars. Since for $v_{max}=2$ the speed of a car is determined by
the states of, at most, two sites in front of it, the minimal block
size to obtain nontrivial results is 3 (the site occupied by a car
plus two sites in front of it). In what follows,
we limit our attention to order-3 local structure approximation.

Using 3-block probabilities, we can write a set of equations
describing the time evolution of these probabilities

\begin{equation} \label{lstexact}
P_{t+1}(b_2b_3b_4)=\sum_{a_i \in \{0,1\}\atop i=0,1,\ldots,6}
w(b_2b_3b_4 | a_0a_1a_2a_3a_4a_5a_6) P_t(a_0a_1a_2a_3a_4a_5a_6),
\end{equation}
where $P_t(b_2b_3b_4)$ is the probability of block $b_2b_3b_4$ at
time $t$,\\
and $w(b_2b_3b_4 | a_0a_1a_2a_3a_4a_5a_6)$ is the
conditional probability that the rule maps the seven-block
$a_0a_1a_2a_3a_4a_5a_6$ into the three-block $b_2b_3b_4$. Letters
$a$ denote the states of lattice sites at time $t$, while $b$
denote states at time $t+1$, so that, for example, $a_3$ is the
state of site $i=3$ at time $t$, and $b_3$ is the state of the
same site at time $t+1$.  Conditional probabilities $w$ can be
easily computed from the definition of the rule, although, since
their number is quite large, we used a computer program for their
determination.

Equation (\ref{lstexact}) is exact. The approximation consists of
expressing the seven-block probabilities in terms of three-block
probabilities using Bayesian extension. That is,
\begin{eqnarray} \label{bayesian}
&&P_t(a_0a_1a_2a_3a_4a_5a_6)=\\ && \mbox{\ \ \ \ \ \ \ \ }
\frac{P_t(a_0a_1a_2)P_t(a_1a_2a_3)P_t(a_2a_3a_4)
P_t(a_3a_4a_5)P_t(a_4a_5a_6)}{ P_t(a_1a_2)P_t(a_2a_3)P_t(a_3a_4)
P_t(a_4a_5)}, \nonumber
\end{eqnarray}
where $P_t(a_ia_{i+1})=P_t(a_ia_{i+1}0)+P_t(a_ia_{i+1}1)$ for
$i=1,\ldots,4$. Equations (\ref{lstexact}) and (\ref{bayesian})
define a dynamical system whose fixed point approximates
three-block probabilities of the limit set of the CA rule. Due to
the nonlinear nature of these equations it is not possible to find the
fixed point analytically. We have determined it numerically.

From the knowledge of three-block probabilities $P_t(a_2a_3a_4)$, the
car density is given by
\begin{equation}
\rho=P_t(1)=P_t(100)+P_t(101)+P_t(110)+P_t(111),
\end{equation}
and the average velocity by
\begin{equation}
\langle v \rangle =\frac{2 (1-p) P_t(100) +
p P_t(100) + (1-p) P_t(101) }{\rho}.
\end{equation}
This last result gives the expression of the order parameter
$m=v_{max}-\langle v\rangle$.

Note that $\rho=P_t(1)$ is a constant of motion for equation
(\ref{lstexact}) whose fixed point depends, therefore,
upon the initial car density
$P_0(1)=P_0(100)+P_0(101)+P_0(110)+P_0(111)$. Consequently, using
order-3 local structure approximation allows to determine the order
parameter $m$ as a function of car density, and the critical exponents.

For most simulations, we used a lattice size equal to 1000 and our
results are average of 1000 runs of 1000 iterations. For
$p=0.0005$ we took a lattice of 10000 sites and averaged 500 runs
of 10000 iterations.

\section{Results}
\begin{figure}
\begin{center}
a)\includegraphics[scale=0.85]{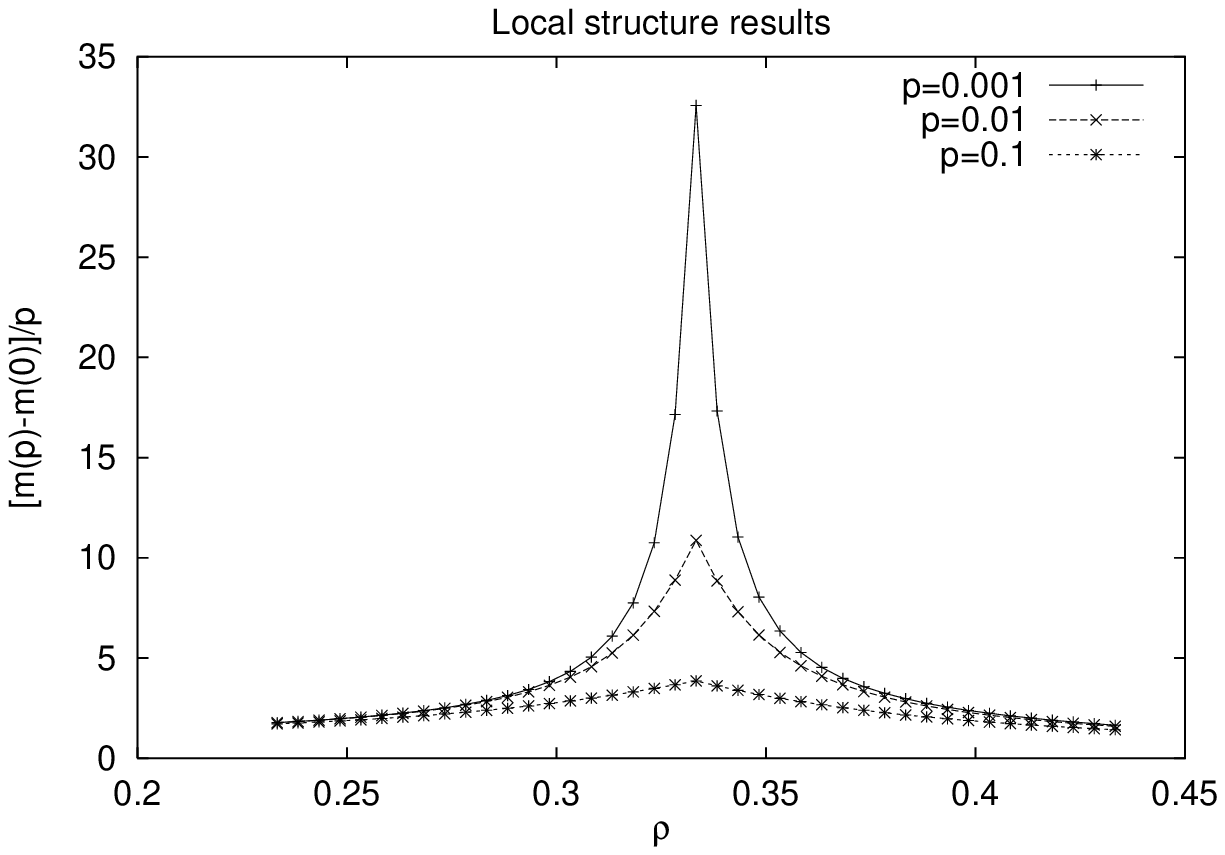}
b)\includegraphics[scale=0.85]{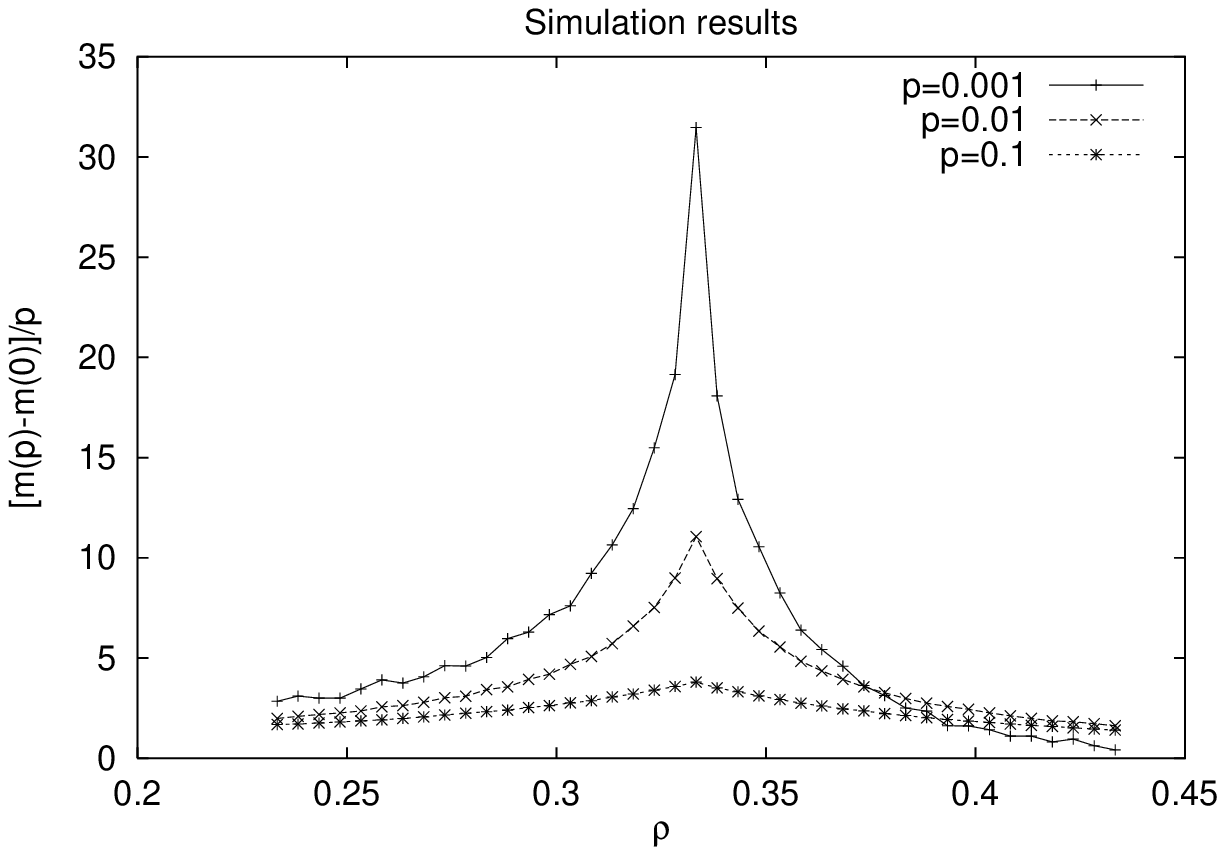} \caption{Divergence of
the susceptibility at the critical point using local structure
approximation (a) and data obtained from simulations (b).
Excellent agreement of simulations and local structure
approximation is clearly visible.}\label{peak}
\end{center}
\end{figure}

\begin{figure}
\begin{center}
a)\includegraphics[scale=0.85]{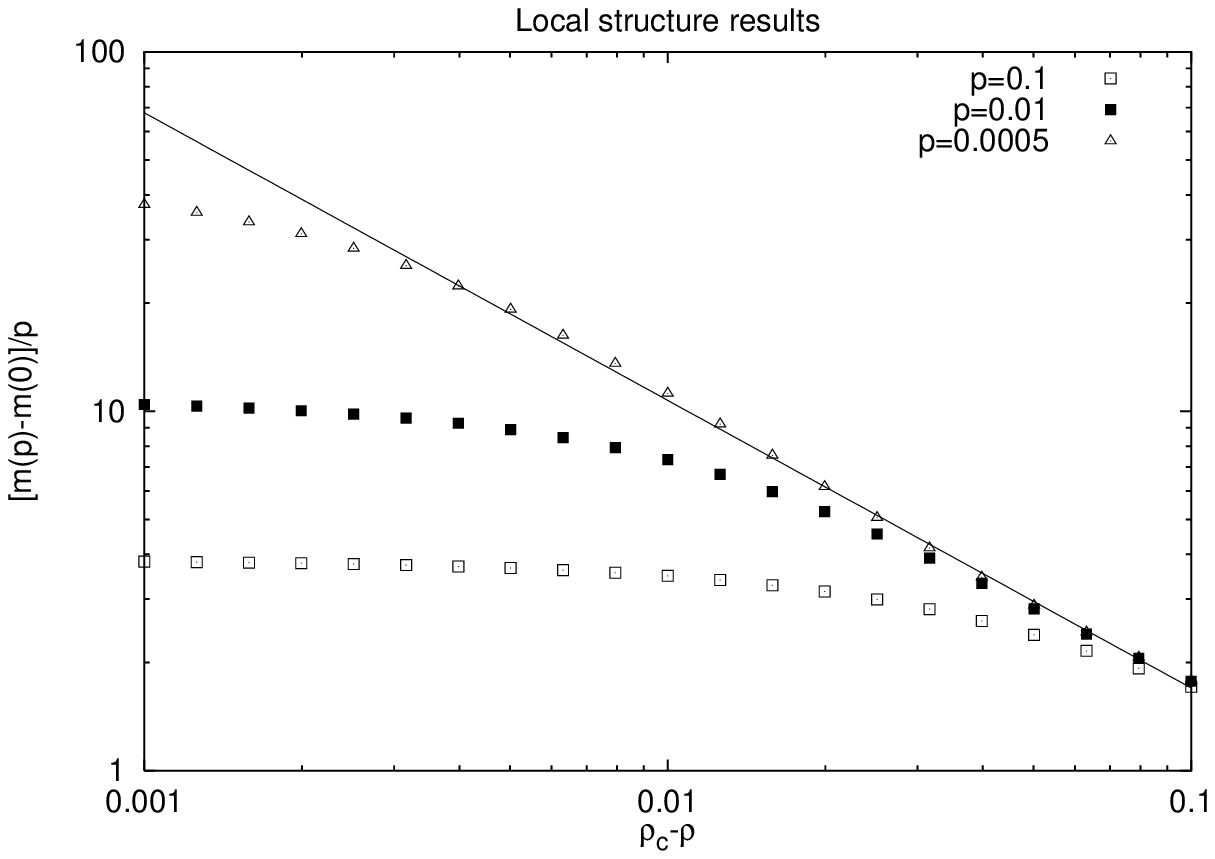}
b)\includegraphics[scale=0.85]{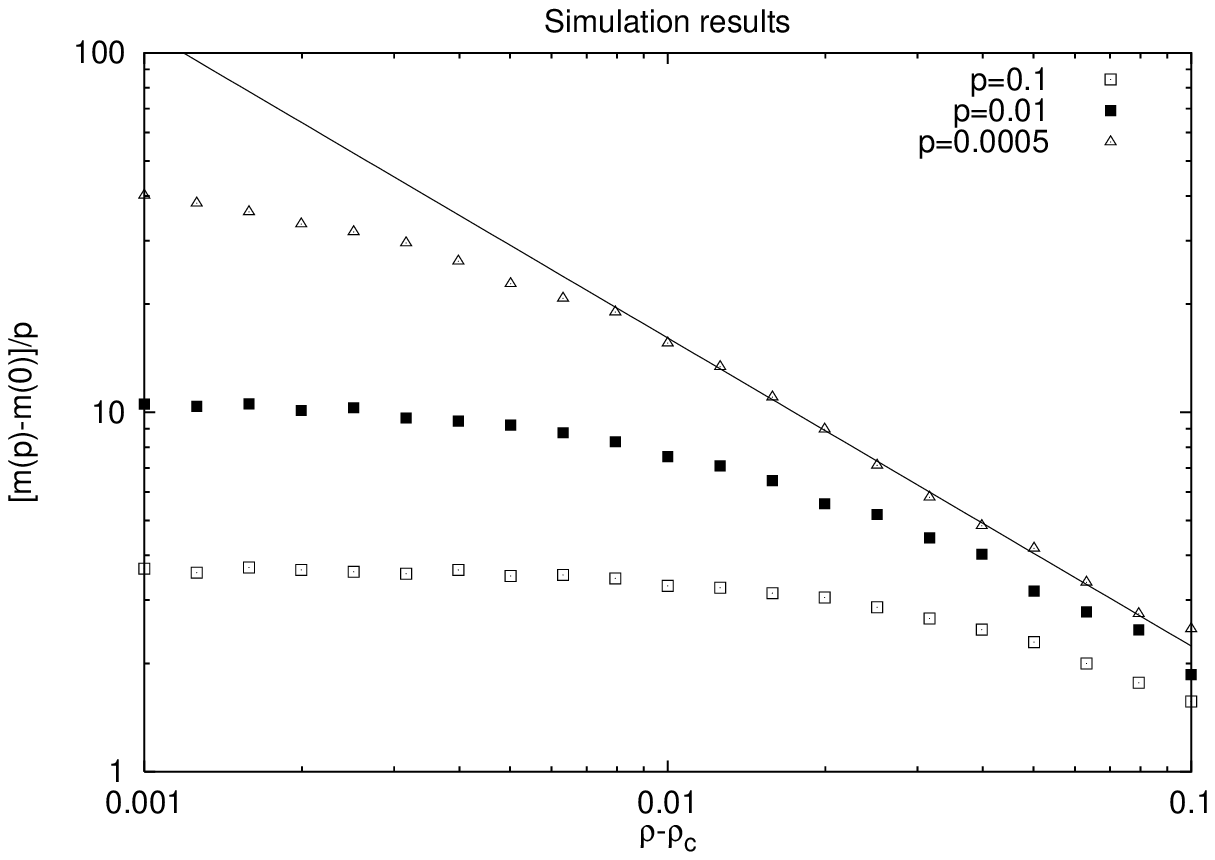} \caption{Plot of
$[m(p)-m(0)]/p$ as a function of $|\rho-\rho_c|$ for $\rho<\rho_c$
using local structure approximation (a) and data obtained from
simulations (b). Straight line represents best fit for $p=0.0005$
and $0.01<\rho_c-\rho<0.1$. We obtained $\gamma$ by performing
such a fit for several different values of $p$ and extrapolation
to $p=0$.}\label{gamma}
\end{center}
\end{figure}

\begin{figure}
\begin{center}
a)\includegraphics[scale=0.85]{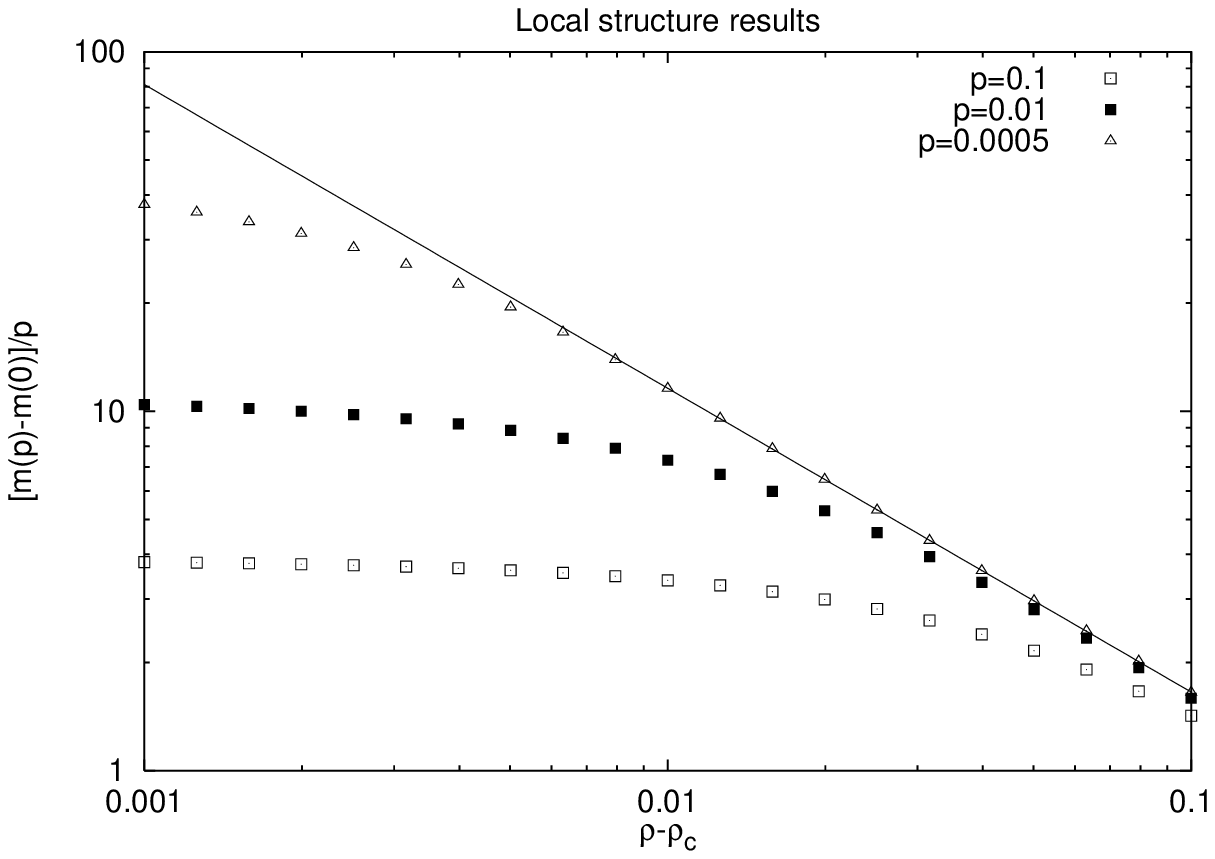}
b)\includegraphics[scale=0.85]{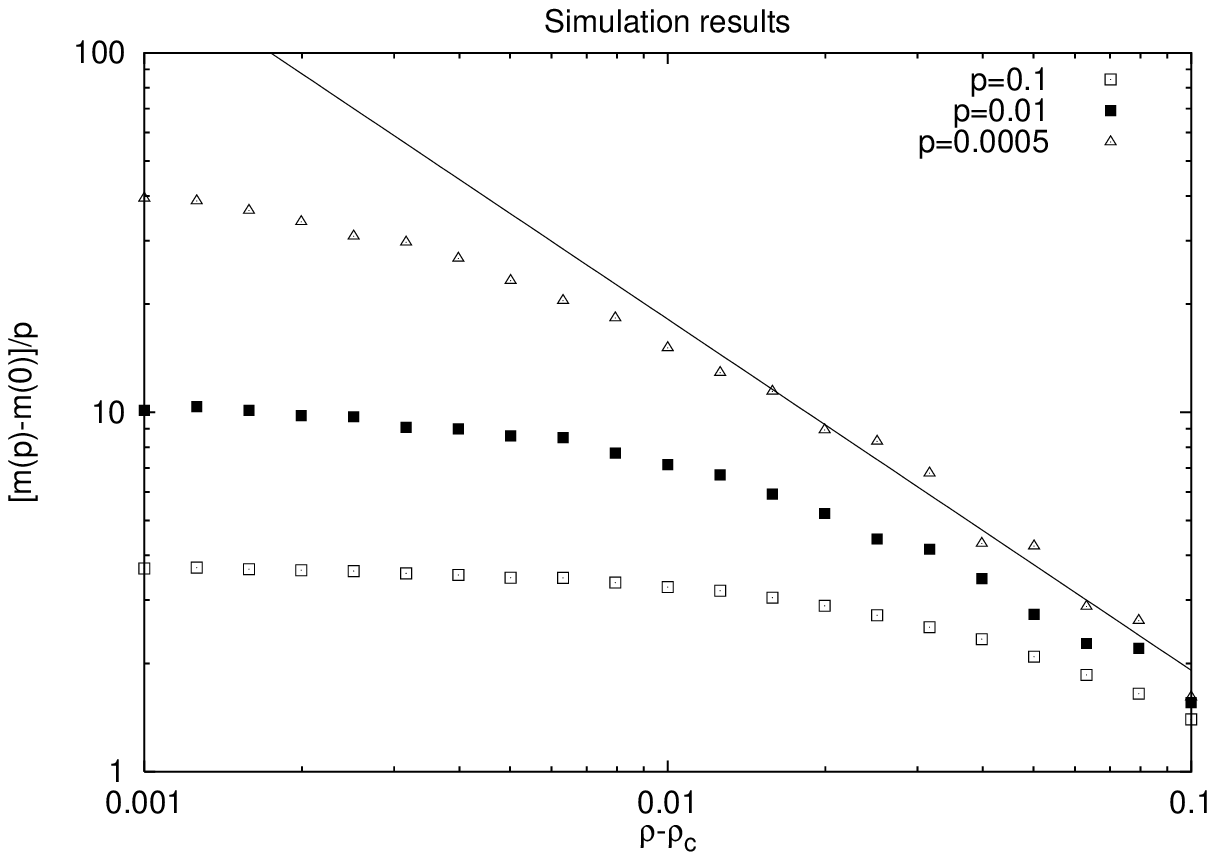} \caption{Plot of
$[m(p)-m(0)]/p$ as a function of $|\rho-\rho_c|$ for $\rho>\rho_c$
using local structure approximation (a) and data obtained from
simulations (b). Straight line represents best fit for $p=0.0005$
and $0.01<\rho-\rho_c<0.1$. We obtained $\gamma^{\prime}$ by
performing such a fit for several different values of $p$ and
extrapolation to $p=0$.}\label{gammap}
\end{center}
\end{figure}

The susceptibility $\chi_{\rho}$ at constant $\rho$, defined by
\begin{equation}
\chi_{\rho} = \lim_{p\to 0}\frac{\partial m}{\partial p},
\end{equation}
cannot be calculated exactly. So we determined its value using
local structure approximation as described in the preceding section
and computer simulations for $v_{\max}=2$. Figures \ref{peak}a and
\ref{peak}b show the results obtained from order-3 local
structure approximation\footnote{If $v_{\max}>2$, one should use
$(v_{\max}+1)^{\rm th}$-order local structure approximation} and
simulations,  respectively. In the limit $p\to 0$, the
susceptibility diverges as $(\rho_c-\rho)^{-\gamma}$ for
$\rho<\rho_c$, and as $(\rho-\rho_c)^{-\gamma'}$ for
$\rho>\rho_c$. Using local structure approximation (Figures
\ref{gamma}a and \ref{gammap}a), we found
$$
\gamma = 0.91\pm0.03\quad\text{and}\quad\gamma' = 0.98\pm0.03
$$
while simulations (Figures \ref{gamma}b and \ref{gammap}b) yield
$$
\gamma = 0.86\pm0.05\quad\text{and}\quad\gamma' = 0.94\pm0.05.
$$

\begin{figure}
\begin{center}
a)\includegraphics[scale=0.85]{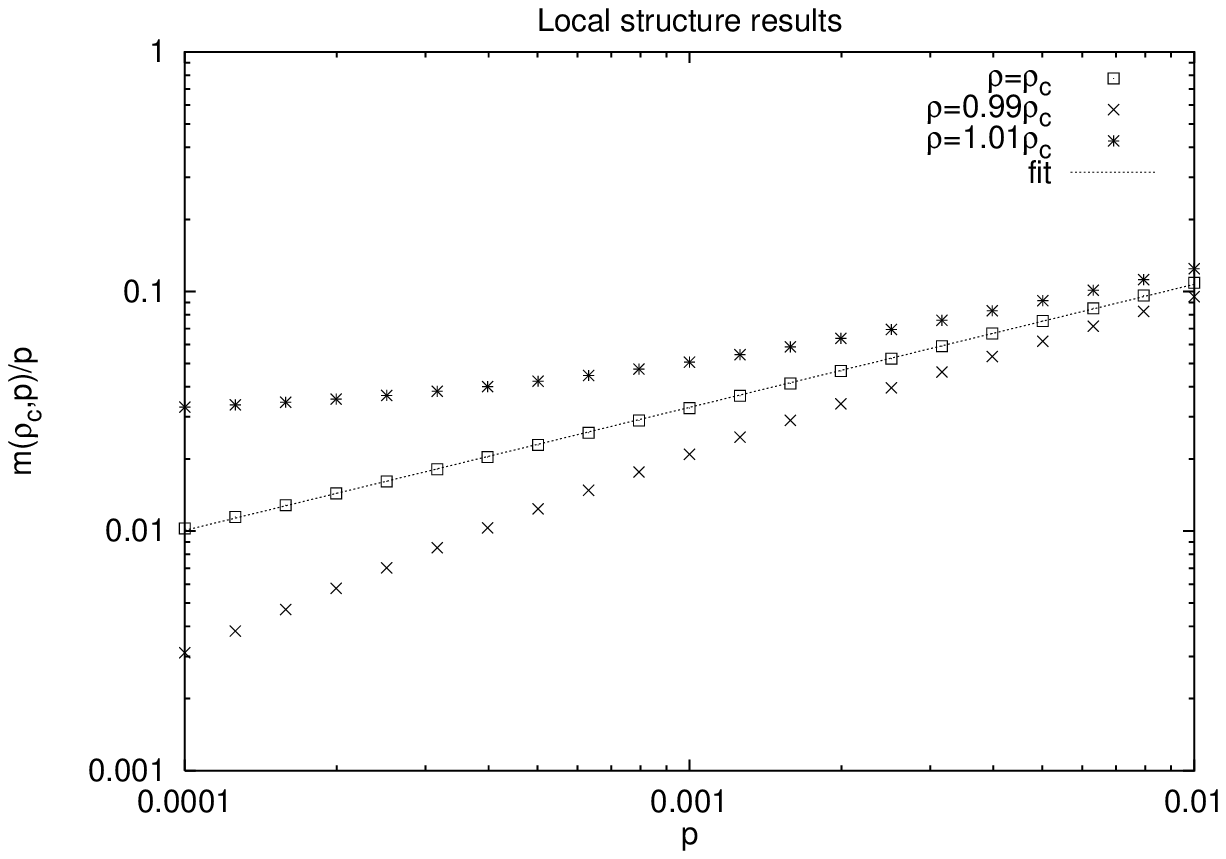}
b)\includegraphics[scale=0.85]{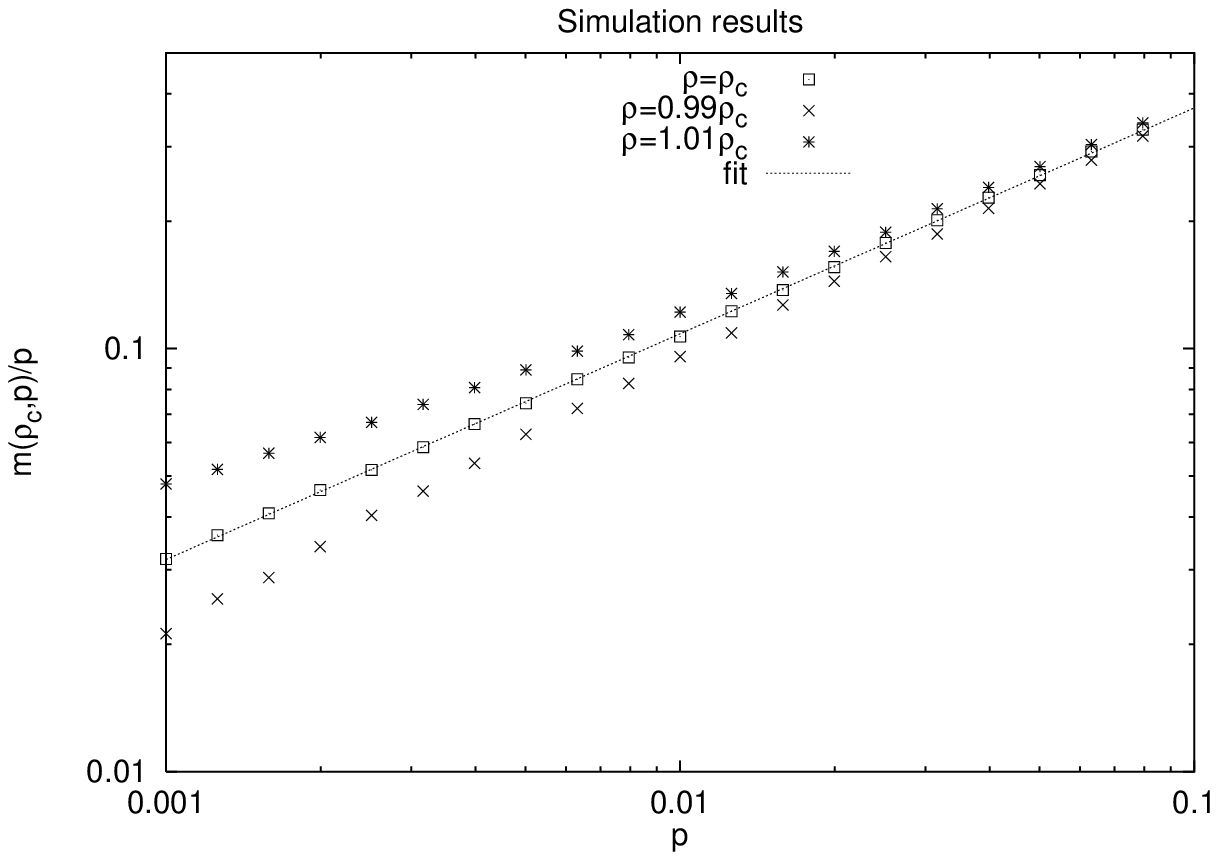} \caption{Order
parameter $m$ at the critical point $\rho=\rho_c$ as a function of
$p$ using local structure approximation (a) and data obtained from
simulations (b). Slope of the best fit line at $\rho=\rho_c$
yields $1/\delta$.}\label{delta}
\end{center}
\end{figure}

Another exponent of interest is $\delta$. It characterizes the
behavior of $m$ as a power of $p$ for $\rho=\rho_c$. Here again we
have determined the value of $$ \lim_{p\to
0}\frac{m(\rho_c,0)-m(\rho_c,p)}{p} $$ using order-3 local
structure approximation and simulations. Our results are
represented in Figure \ref{delta}, and the values of $\delta$,
obtained using order-3 local structure approximation and computer
simulations, are, respectively, given by
$$
1/\delta = 0.51\pm0.01\quad\text{and}\quad1/\delta = 0.53\pm0.02
$$

In is interesting to note that the values
$$
\beta = 1, \quad\gamma\approx 1, \quad\delta\approx 2
$$
obtained for the critical exponents are found in equilibrium
statistical physics in the case of second-order phase transitions
characterized by nonnegative order parameters above the upper
critical dimensionality.

\begin{figure}
\begin{center}
a)\includegraphics[scale=0.85]{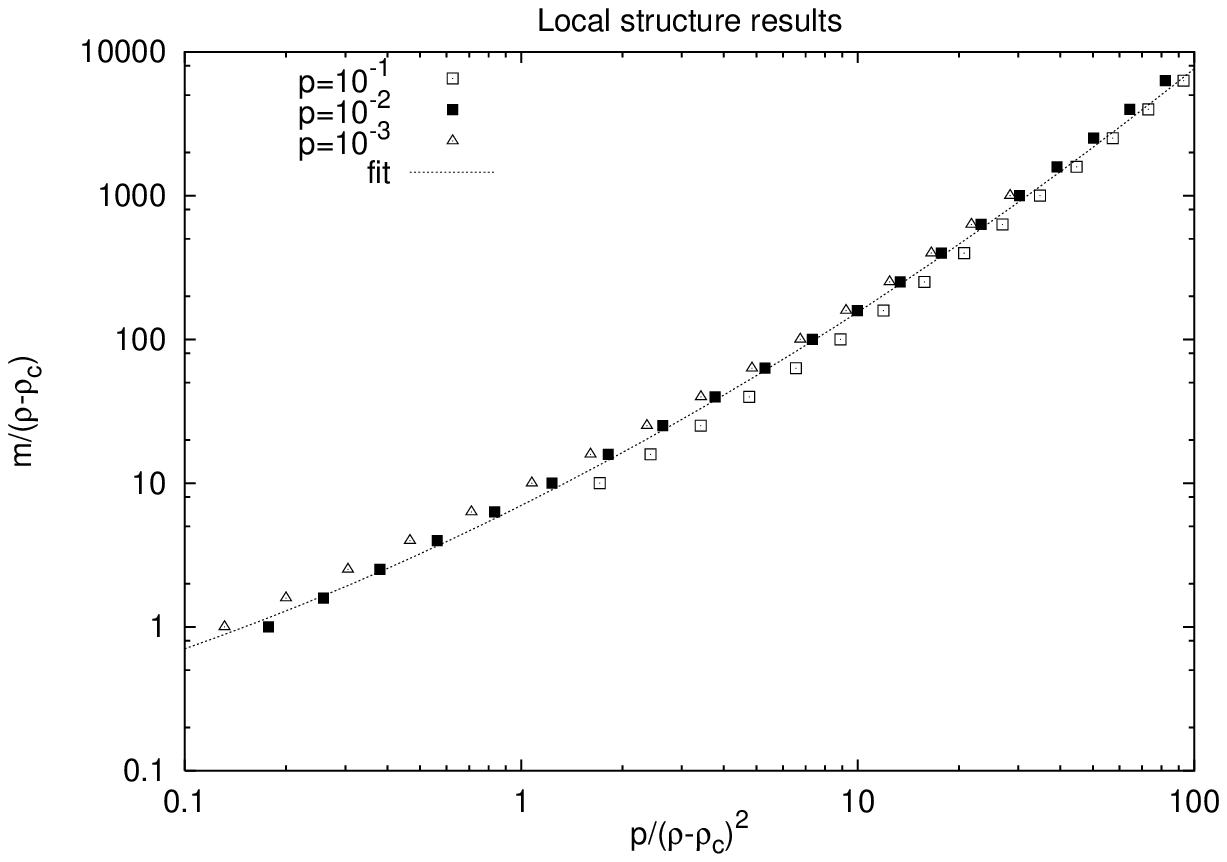}
b)\includegraphics[scale=0.85]{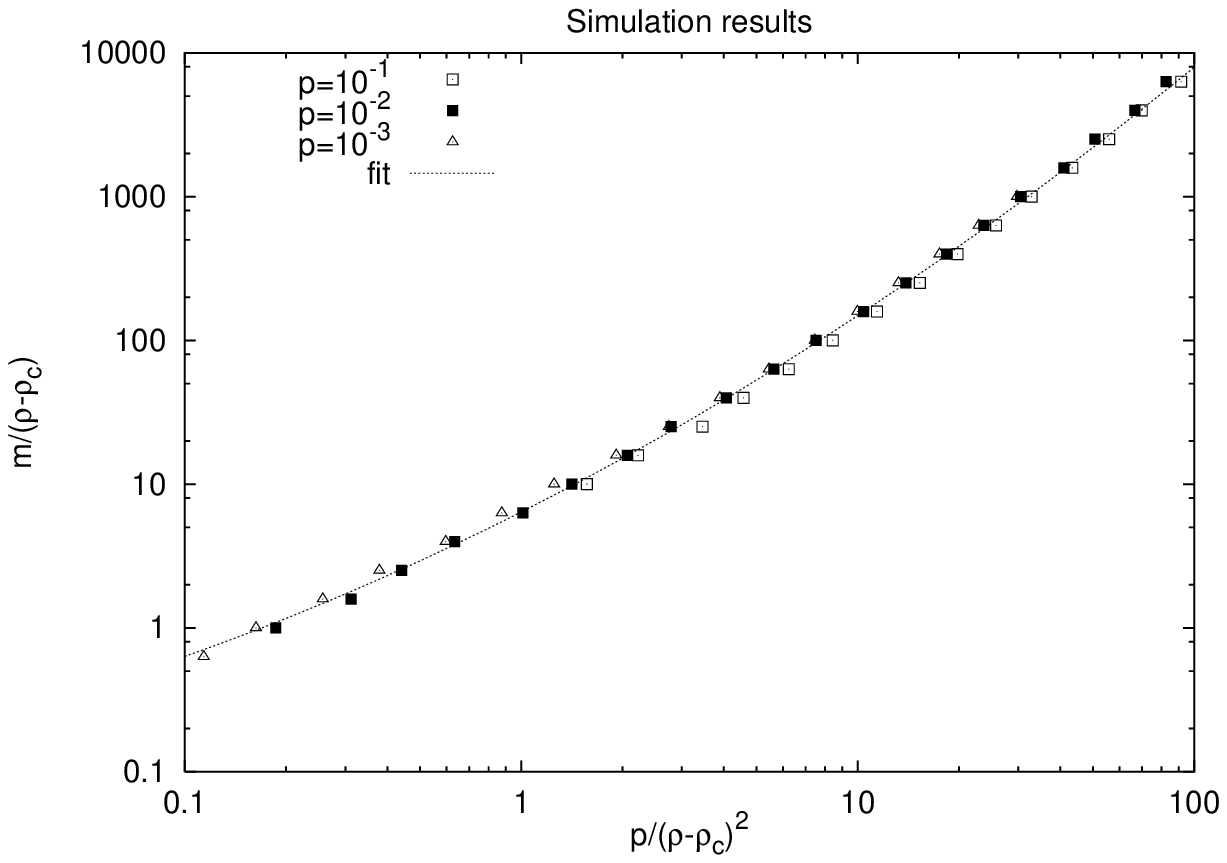} \caption{Illustration
of the existence of universal scaling function for $\rho<\rho_c$
using local structure approximation (a) and data obtained from
simulations (b).}\label{scal}
\end{center}
\end{figure}

\begin{figure}
\begin{center}
a)\includegraphics[scale=0.85]{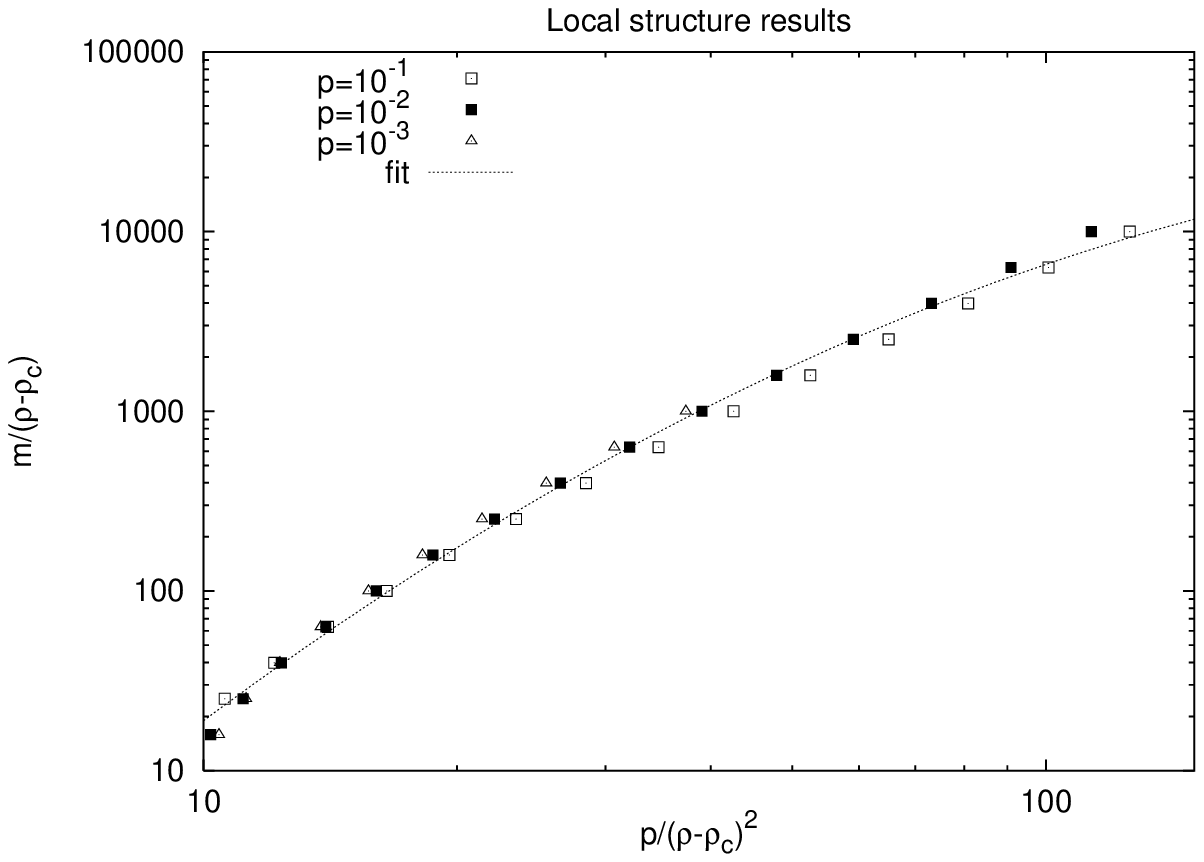}
b)\includegraphics[scale=0.85]{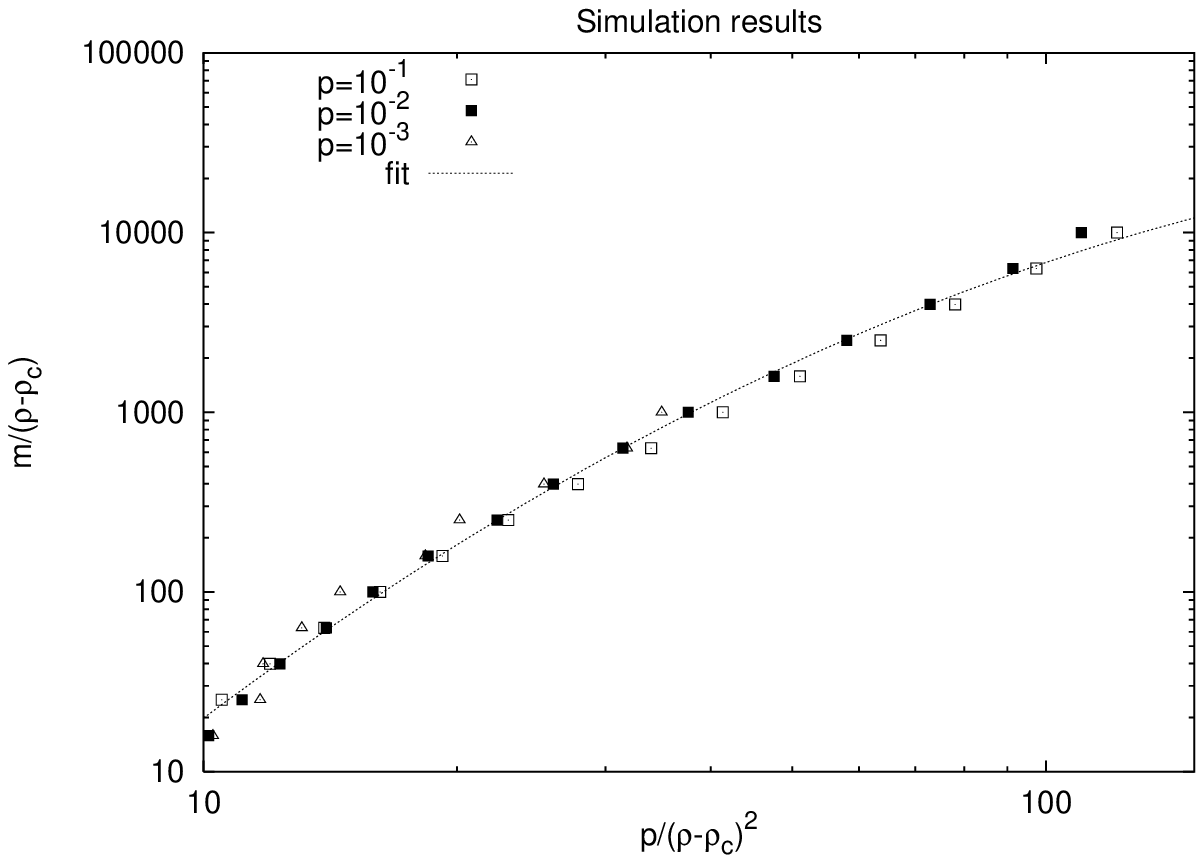} \caption{Illustration
of the existence of universal scaling function for $\rho>\rho_c$
using local structure approximation (a) and data obtained from
simulations (b).}\label{scalp}
\end{center}
\end{figure}

Close to the phase transition point, critical exponents obey
scaling relations. If we assume that, in the vicinity of the
critical point $(\rho=\rho_c, p=0)$, the order parameter $m$ is a
generalized homogeneous function of $\rho-\rho_c$ and $p$ of the
form
\begin{equation}
m = |\rho-\rho_c|^\beta
f\left(\frac{p}{|\rho-\rho_c|^{\beta\delta}}\right),
\end{equation}
where the function $f$ is such that $f(0)\neq 0$, then,
differentiating $f$ with respect to $p$ and taking the limit
$p\to 0$, we readily obtain
\begin{equation}
\gamma = \gamma' = (\delta-1)\beta,
\end{equation}
which is verified by our numerical values obtained either using
order-3 local structure approximation or computer simulations.
Figures \ref{scal} and \ref{scalp},  clearly confirm the existence
of a universal scaling function.

\section{Conclusion}
In this short note, we have derived the critical behavior of a CA
traffic flow model using an order parameter breaking the symmetry
of the jam-free phase. Random braking, which is thought to be an
essential ingredient of any CA traffic flow model, appears to be
the symmetry-breaking field conjugate to the order parameter. For
$v_{\max}=2$, we have determined the critical exponents $\beta$,
$\gamma$ and $\delta$ using order-3 local structure approximation
and computer simulations. These critical exponents satisfy a
scaling relation which can be derived assuming that the order
parameter is a generalized homogeneous function of $|\rho-\rho_c|$
and $p$ in the vicinity of the phase transition point.

\section{Acknowledgements}
One of the authors (H. F.) acknowledges financial support from the Natural
Sciences and Engineering Research Council (NSERC) of Canada. He also wishes
to thank The Fields Institute for Research in Mathematical Sciences
and the Department of Mathematics and Statistics of University of Guelph
for generous hospitality.

%%%%%%%%%%%%%%%%%%%%%%%%%%%%%%%%%%%%%%%%%%%%%
%%%%%%%%%%%%Figures come here%%%%%%%%%%%%%%%%
%%%%%%%%%%%%%%%%%%%%%%%%%%%%%%%%%%%%%%%%%%%%%

\begin{thebibliography}{0000}

\bibitem{NaSch} K. Nagel and M. Schreckenberg,
J. Physique (France) {\bf I2} 2221 (1992).

\bibitem{Schad} A. Schadschneider, preprint
\texttt{cond-mat/9902170}.

\bibitem{ESSS} B. Eisenbl\"atter, L. Santen, A. Schadschneider,
and M. Schreckenberg, Phys. Rev. E {\bf 57} 1309 (1998).

\bibitem{LSU} S. L\"ubeck, M. Schreckenberg, and K. D. Usadel,
Phys. Rev. E {\bf 57} 1171 (1998).

\bibitem{FK} M. Fukui and Y. Ishibashi, J. Phys. Soc. Japan
{\bf 65} 2345 (1996).

\bibitem{BNR} N. Boccara, J. Nasser and M. Roger, Phys. Rev. A
{\bf 44} (2) 866 (1991).

\bibitem{BF1} N. Boccara and H. Fuk\'s, J. Phys. A: Math. Gen.
{\bf 31} 6007 (1998) and \texttt{adap-org/9712003}.

\bibitem{BF2} N. Boccara and H. Fuk\'s, preprint
\texttt{adap-org/9905004}.

\bibitem{HF} H. Fuk\'s, Phys. Rev. E  {\bf 60} 197 (1999) and
\texttt{comp-gas/9902001}.

\bibitem{GVK} H. A. Gutowitz, J. D. Victor, and B. W. Knight, Physica D
{\bf 28} 18 (1987).

\end{thebibliography}
\end{document}